\documentclass[submission,copyright,creativecommons]{eptcs}
\providecommand{\event}{iFMCloud 2016} 
\usepackage[english]{babel}
\usepackage[utf8]{inputenc}
\usepackage[T1]{fontenc}
\usepackage{listings}[2002/04/01]  
\usepackage{courier}
 \usepackage{graphicx}
\usepackage{amssymb} 
\usepackage{stmaryrd} 
\usepackage{breakurl}             
\usepackage{underscore}           
\usepackage[dvipsnames]{xcolor,colortbl}
\usepackage{amsmath,cite,xspace,latexsym}
\usepackage{hyperref}
\usepackage{tikz}

\newcommand{\ABS}{ABS\xspace}
\newcommand{\rtabs}{Real Time ABS\xspace}
\newcommand{\abscloud}{ABS Cloud API\xspace}

\definecolor{Green}{rgb}{0,0.8,0}
\definecolor{gray20}{gray}{0.95}
 \def\myttsize{\fontsize{8}{8}}
\lstdefinelanguage{ABS}{keywords=
{null,this,thisDC,dyndelta,new,data,type,def,case,of,cog,class,interface,extends,implements,if,else,await,get,total,load,transfer,movecogto,Fut,return,skip,while,module,duration,now,import, export, uses, from, destiny, suspend,delta,adds,modifies,removes,original,productline,features,core,corefeatures,optionalfeatures,after,when,product,hasAttribute,hasMethod,root,extension,group,allof,oneof,require,exclude,original,ifin,ifout,opt}, sensitive=true, comment=[l]{//}, morecomment=[s]{/*}{*/}, morestring=[b]"}

\lstdefinestyle{absstyle}{
language=ABS,columns=fullflexible,
 		   mathescape=true,%
 		   showstringspaces=false,%
keywordstyle=\bf\ttfamily,
commentstyle=\sl\ttfamily,%
basicstyle=\footnotesize\ttfamily,
inputencoding=latin1, 
extendedchars,xleftmargin=2em
}

\newcommand{\Abs}[1]{\lstinline[language=ABS,columns=fullflexible,mathescape=true,inputencoding=latin1,extendedchars,keywordstyle=\bf\ttfamily,basicstyle=\ttfamily]!#1!}

\newenvironment{absinline}[1][]{%
\lstset{language=ABS,
		 showstringspaces=false,
		 columns=fullflexible,mathescape=true,%
         keywordstyle=\bf\ttfamily,commentstyle=\sl\ttfamily,%
         basicstyle=\normalsize\ttfamily,inputencoding=latin1, 
         extendedchars,xleftmargin=2em,#1}}
{}

\lstnewenvironment{absexample}{
~\\
\noindent 
\small{\textcolor{Green!80!black}{\bfseries\sffamily{Example:}}}\vspace{-4pt}
\lstset{style=absstyle,
basicstyle=\myttsize\ttfamily,
framerule=1pt,
backgroundcolor=\color{Green!5},
rulecolor=\color{Green!80!black},
frame=tblr,
xleftmargin=4pt,
}}
{}

\lstnewenvironment{absexamplen}[1][]{
\lstset{style=absstyle,
basicstyle=\myttsize\ttfamily,
framerule=1pt,tabsize=2,
backgroundcolor=\color{Green!5},
rulecolor=\color{Green!80!black},
frame=tblr,
captionpos=b,
xleftmargin=4pt,#1
}}
{}

\title{Modeling Deployment Decisions for Elastic Services with ABS\thanks{This work was done in the context of the EU project
    FP7-610582 \emph{ENVISAGE: Engineering Virtualized Services} (\texttt{http://www.envisage-project.eu}).}}
\author{Einar Broch Johnsen, Ka I Pun, S.\ Lizeth Tapia Tarifa
\institute{Department of Informatics, University of Oslo, Norway\\
\{\url{einarj,violet,sltarifa}\}\url{@ifi.uio.no}}}

\def\titlerunning{Modeling Deployment Decisions for Elastic Services with ABS}
\def\authorrunning{Einar Broch Johnsen, Ka I Pun \& S.\ Lizeth Tapia Tarifa}

\begin{document}
\maketitle

\begin{abstract}
  The use of cloud technology can offer significant savings for the
  deployment of services, provided that the service is able to make
  efficient use of the available virtual resources to meet
  service-level requirements. To avoid software designs that scale
  poorly, it is important to make deployment decisions for the service
  at design time, early in the development of the service itself. \ABS
  offers a formal, model-based approach which integrates the design of
  services with the modeling of deployment decisions. In this paper,
  we illustrate the main concepts of this approach by modeling a
  scalable pool of workers with an auto-scaling strategy and by using
  the model to compare deployment decisions with respect to client
  traffic with peak loads.
\end{abstract}

\section{Introduction}

Insufficient scalability and bad resource management of software
services can easily consume any potential savings from cloud
deployment. Failed service-level agreements (SLAs) cause penalties for
the provider, while oversized SLAs waste resources on the customer's
side.  IBM Systems Sciences Institute estimates that a defect which
costs one unit to fix in design, costs 15 units to fix in testing
(system/acceptance) and 100 units or more to fix in production
\cite{BoehmPapaccio88}; this cost estimation does not even consider
the \emph{impact cost} due to, for example, delayed time to market,
lost revenue, lost customers, and bad public relations.

Deployment on the cloud gives software designers far reaching control
over the resource parameters of the execution environment, such as the
number and kind of processors, the amount of memory and storage
capacity, and the bandwidth.  In this context, designers can also
control their software's trade-offs between the incurred cost and the
delivered quality-of-service. SLA-aware services, which are
\emph{designed for scalability}, can even change these parameters
dynamically, at runtime, to meet their service contracts.

The formally defined \emph{\bfseries A}bstract \emph{\bfseries
  B}ehavioral \emph{\bfseries S}pecification language
\ABS~\cite{johnsen10fmco} realizes a separation of concerns between
the \emph{cost} of execution and the \emph{capacity} of dynamically
provisioned cloud resources~\cite{johnsen15jlamp}. Models are
executable; a simulation tool for \ABS supports rapid prototyping and
visualization.  The use of languages such as \ABS enables developers
to shift deployment decisions from late in the software engineering
process to become an integral part of software design
\cite{HaehnleJohnsen14}.  \ABS permits to design and validate these
services by connecting executable models to quality of service
requirements, using a \emph{Cloud API} to interface with an
abstraction of the cloud provisioning.  The modeling approach and
analyses developed for \ABS have been successfully applied in an
industrial context to SDL Fredhopper's eCommerce
Optimization~\cite{ABHJSTW13} and to Apache's Hadoop
YARN~\cite{lin16fase}.

In this paper, we illustrate the use of \ABS to make deployment
decisions at the modeling level for a so-called \emph{hot pool}
\cite{Fehling:CCP2014}: a local scaling point in a service with a load
balancer distributing jobs to workers. Hot pools were originally
introduced for resilience, but represent a viable approach for
fine-grained scaling of services on the cloud. We model different
deployment scenarios for the hot pool, varying in the numbers of
workers, and show how to model a simple autoscaler for the hot
pool. We use simulations to compare performance and resource usage for
the hot pool using different deployment scenarios. We refer to
\cite{johnsen15jlamp,ABHJSTW13} for related work on the \ABS modeling
language.


\section{The ABS Language}

\ABS is a modeling language for the development of executable distributed and deployed object-oriented models.  The main characteristics of \ABS can be listed as follows: 
\begin{enumerate}
\item it has a formal syntax and semantics;
\item it has a clean integration of concurrency and object orientation based on concurrent object groups (COGs) \cite{schaefer10ecoop,johnsen10fmco};
\item it permits synchronous as well as asynchronous communication \cite{johnsen07sosym, deboer07esop};
\item it offers a wide variety of complementary modeling alternatives that integrates a functional layer with algebraic datatypes and functional programming, an imperative layer~\cite{haahnle13, ABSTutorial2013, johnsen10fmco} with COGs and asynchronous communication, it allows the modeling of real-time behavior~\cite{bjork13isse};
\item compared to object-oriented programming languages, it abstracts from low-level implementation choices for data structures, and compared to design-oriented languages like UML diagrams, it is executable and models the control flow of systems;
\item it supports deployment modeling by means of a separation of concerns between the resource costs of executions and the resource capacities of (virtual) locations. Deployment decisions can be made inside the models~\cite{johnsen15jlamp}, using a Cloud API to interact with the cloud provisioning layer~\cite{johnsen12};
\end{enumerate}

\emph{The functional layer} of \ABS is used to model computations on the internal data of objects. It allows designers to abstract from the implementation details of imperative data structures at an early stage in the software design.  The functional layer combines a language for parametric algebraic data types (ADTs) and a simple functional language with case distinction and pattern matching. \ABS includes a library with predefined datatypes such as \Abs{Bool}, \Abs{Int}, \Abs{String}, \Abs{Rat}, \Abs{Unit}, etc. It also has parametric datatypes such as lists, sets and maps.  All other types and functions are user-defined.

\emph{The imperative layer} of \ABS allows designers to express communication and synchronization between concurrent objects. In the imperative layer, processes are encapsulated within COGs~\cite{schaefer10ecoop,johnsen10fmco}, the processes are created automatically at method call reception and terminated after the method call execution is finished.  \ABS combines active (with a \Abs{run} method which is automatically activated) and reactive behavior of objects. \ABS is based on cooperative scheduling: Inside COGs processes may suspend at explicitly defined scheduling points, at which point control may be transferred to another process.
Suspension allows other pending processes to be activated. However, the suspending process does not signal any particular process, instead the selection of the new process is left to the scheduler. In between these explicit scheduling points, only one process is active inside a COG, which means that race conditions are avoided. \emph{\rtabs}~\cite{bjork13isse} extends \ABS with support for the modeling and manipulation of dense time. This extension allows to represent execution time  inside methods. The local passage of time is expressed in   terms of a statement called \Abs{duration} (as in, e.g., UPPAAL \cite{larsen97sttt}).
To express dense time, we consider a  model represented by two
types \Abs{Time} and \Abs{Duration}.  Time values capture points in
time as reflected on a global clock during execution. In contrast,
finite durations reflect the passage of time. 

The \abscloud provides an interface to model cloud infrastructure in \ABS~\cite{johnsen12}. The \abscloud supports the dynamic acquisition and release of virtual machines with resources.  Virtual machines in \ABS are modeled using \emph{deployment components}~\cite{johnsen15jlamp}. A deployment component is a modeling abstraction which captures locations offering resources to computations.  The language also supports cost annotations to model resource consumption. The combination of deployment components with resource computations and cost annotations  allows modeling   implicit passage of time. 
In this case, 
time can be observed by measuring the executing model, and monitoring the response time of a system.   In this paper we will model so-called elastic computing resources, where the computation \emph{speed} of virtual machines is determined by the amount of elastic computing resources allocated to these machines per discrete time interval (referred as \emph{time interval} in the rest of the paper).  \emph{The \abscloud} includes methods for launching and shutting down virtual machines. In the implementation, this is done by creating deployment components on which an application manager can deploy objects. In addition, the \abscloud keeps track of the accumulated costs incurred by job execution on the cloud.  \ABS is supported by a range of analysis tools (see, e.g., \cite{ABHJSTW13}); for the analysis results in this paper, we are using the simulation tool which generates Erlang code.


\section{Modeling a Scalable  Service in ABS}

\begin{figure}[t!]
\centering
  \includegraphics[width=0.95\linewidth]{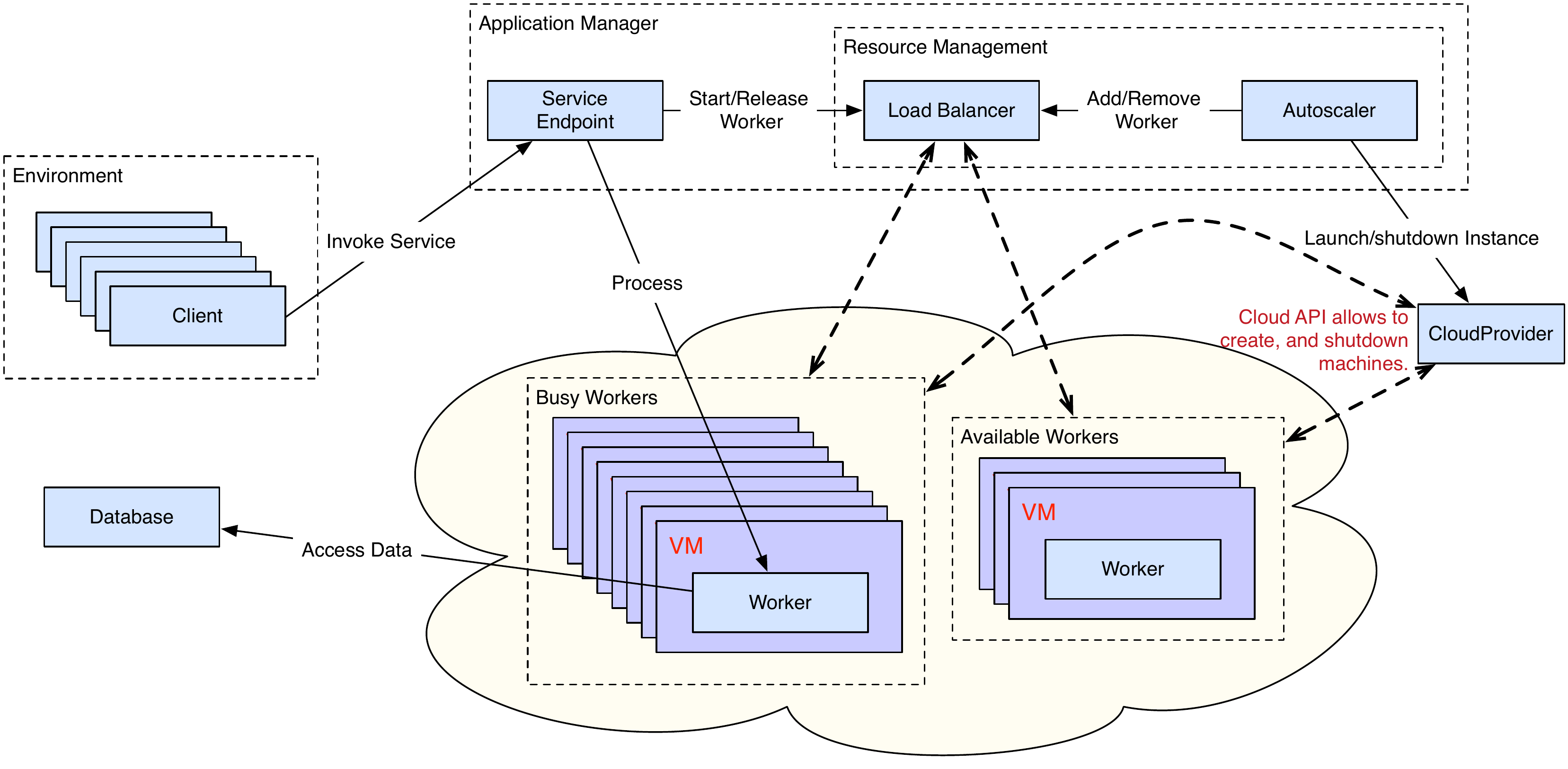}
  \caption{\label{fig:architecture} An architecture for a service with a resource-aware hot pool of workers.}
\end{figure}

Let us consider an example which models a service that executes heavy
and parallel computations on the cloud.  Figure~\ref{fig:architecture}
depicts such an architecture. In this architecture, \emph{clients},
which want to access the service, call a service endpoint. The
\emph{service endpoint} communicates with a load balancer to get
workers that will process the jobs requested by the clients. Each
\emph{worker} is able to process one call of the service at a time,
and for that it needs to access a shared \emph{database}.  The
\emph{load balancer} keeps a list of workers which are currently
processing jobs and a list of workers which are available for new
jobs, and distributes jobs among the workers.  In
Section~\ref{sec:dynamicSys}, we will introduce an autoscaler.  The
\emph{autoscaler} will increase or decrease the number of workers as
needed. For this particular scenario, we will deploy the workers on
the cloud, and each worker will have its own dedicated virtual
machine. Since our concern is the deployment strategy for the hot
pool, we will abstract from the deployment of the other parts of the
system. In \ABS any object which is not explicitly deployed on a
deployment component, will run in an environment with unbounded
available resources.

  \begin{figure}[h!]
\begin{absexamplen}
class ClosedClient (SE ep, Int cycle, Rat cost, Int nbrOfJobs) implements Client {
    Int jobcount = 0;
    Unit run() {
        await duration(cycle, cycle);
        Bool result =  await ep!invokeService(cost); jobcount = jobcount + 1; 
        if (jobcount < nbrOfJobs) { this!run(); }
     }
 }
 
class OpenClient (SE ep, Int cycle, Rat cost, Int nbrOfJobs) implements Client {
    Int jobcount = 0;
    Unit run() { 
        Fut<Bool> fresult =  ep!invokeService(cost); jobcount = jobcount + 1; 
        await duration(cycle, cycle);
        if (jobcount < nbrOfJobs) { this!run(); } await fresult?;  Bool result = fresult.get;
    }
}
\end{absexamplen}
\caption{\label{fig:closeclient} An \ABS model of closed and
  open clients. The former waits for a reply before it sends a new
  invocation, the latter sends a new invocation after a fixed period of
  time.}
\end{figure}

\subsection{Modeling the Environment}
\label{sec:environment}
In our example, the environment consists of clients calling the
service.  We are going to model two kind of clients:
\Abs{ClosedClient} and \Abs{OpenClient}.  Figure~\ref{fig:closeclient}
contains the \ABS model of these clients. The clients' behavior is
captured in the \Abs{run} method. The closed client does not flood the
system. In each iteration this client waits for a cycle to pass, after
that it sends an invocation of the service and waits for the
result. The iteration will finish once the client has sent the desired
number of jobs.  The open client may flood the system. In each cycle
this client sends an invocation without waiting for the result. If the
cycle is very short, the system will receive a sudden burst of
requests.

\begin{figure}[h!]
\centering
  \includegraphics[width=0.7\linewidth]{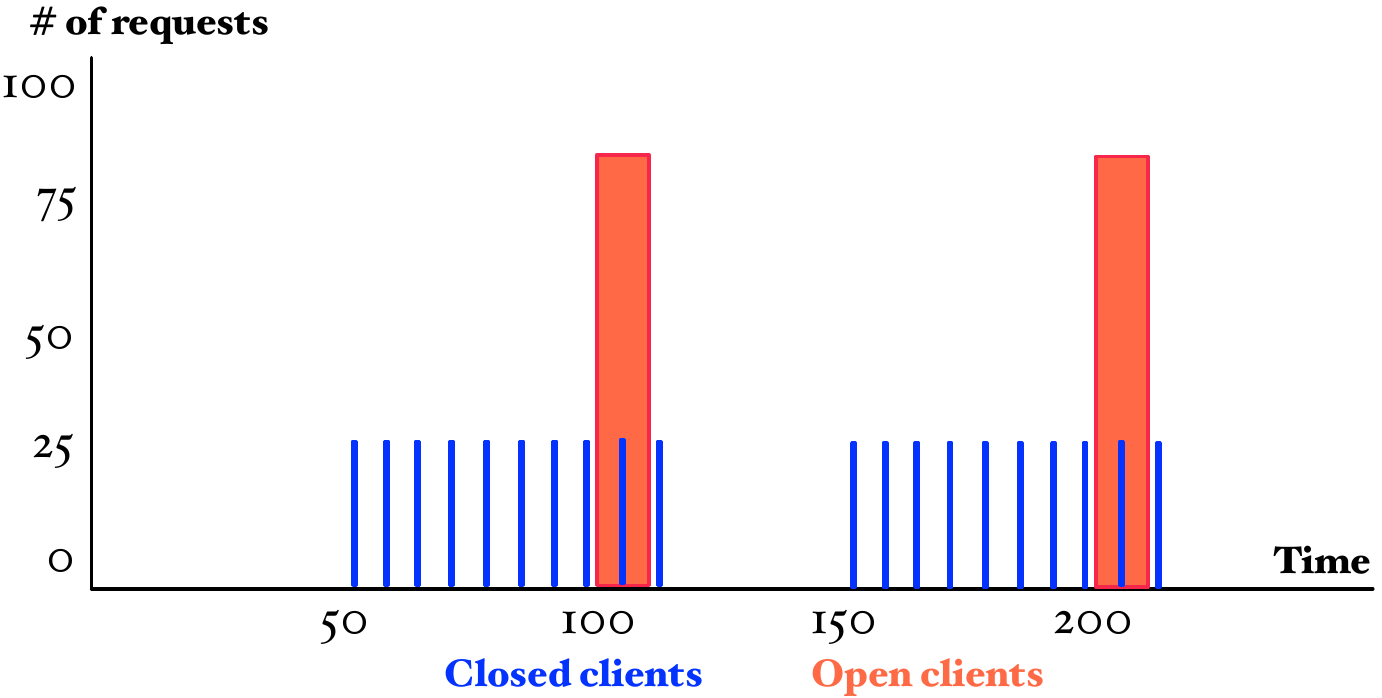}
  \caption{\label{fig:workload} A workload pattern in terms of closed and open clients.}
\end{figure}

Figure~\ref{fig:workload} suggests a workload scenario which is
implemented in Figure~\ref{fig:workload2} using the client behaviors
described above. This workload creates 30 closed clients at time 50
and at time 150 and 80 open clients at time 100 and at time 200. Each
client will send 10 requests, where each request has an average resource cost
of 81, specified in the parameter \Abs{taskCost}. The closed client
has a cycle of 5 time units while the open client has a cycle of one
time unit, creating peaks of requests. We will use this workload to
test the QoS and the accumulated billing cost of the different
deployment scenarios for the system.

\begin{figure}[h!]
\begin{absexamplen}
...
await duration(50,50);  nClts = nClosedClients;
while (nClts > 0) { new ClosedClient(endpoint, 5,taskCost, nbrOfJobs); nClts = nClts - 1;}    
await duration(50,50); nClts = nOpenClients;
while (nClts > 0) { new OpenClient(endpoint, 1, taskCost, nbrOfJobs); nClts = nClts - 1;}
await duration(50,50);  nClts = nClosedClients;
while (nClts > 0) { new ClosedClient(endpoint, 5,taskCost, nbrOfJobs); nClts = nClts - 1;}    
await duration(50,50); nClts = nOpenClients;
while (nClts > 0) { new OpenClient(endpoint, 1, taskCost, nbrOfJobs); nClts = nClts - 1;}
...
  \end{absexamplen}
  \caption{\label{fig:workload2} Implementing the workload pattern in \ABS.}
\end{figure}

\subsection{Modeling the Service with Static Deployment}

In this section we model an initial version of the service. In this case, the application manager consists of the load balancer and the service
endpoint, where the load balancer is statically configured with a fixed
number of workers.

Figure~\ref{fig:worker} contains the \ABS model of a worker which
processes jobs.  The \emph{worker} is modeled by the class
\Abs{WorkerObject} with two methods \Abs{process} and \Abs{getDC}.
The implementation of method \Abs{process} abstracts from the
functional behavior by adding a cost annotation, but captures the
fact that the task needs a transaction to the database, it also takes
into account the time needed to make this transaction. Note that this
implementation can be refined with more explicit functionality and
more fine-grained cost annotations.  The rest of the method
implementation checks whether or not the method call has kept its
deadline. Since we want to check the QoS, we model hard deadlines
instead of soft deadlines.  The method \Abs{getDC} returns the
identifier of the virtual machine where the worker is deployed, using
the \ABS keyword \Abs{thisDC()}.

   \begin{figure}[h!]
\begin{absexamplen}
interface Worker {
    Bool process(Rat taskCost,  Time started, Duration deadline);
    DC getDC();
}

class WorkerObject(Database db) implements Worker {
  
  Bool process(Rat taskCost, Time started, Duration deadline) {
    [Cost: taskCost] skip;
    Duration remainingTime = subtractFromDuration(deadline, timeDifference(now(),started));
    Bool success = await db!accessData(remainingTime);  
    return success;
  }
  
  DC getDC(){ return thisDC();}
}
\end{absexamplen}
  \caption{\label{fig:worker} \ABS model of the worker.}
\end{figure}

Figure~\ref{fig:manager} contains part of the \ABS model of a load
balancer with a round robin scheduling policy.  The
\Abs{RoundRobinLoadBalancer} implements the interface
\Abs{LoadBalancer}. This class keeps two lists, one with the
identifiers of workers which are in use (or busy) and one with the identifiers
of workers which are available. The class also has methods to add and
remove workers to the lists, and to move workers from one list to the
other. Method \Abs{addWorker} adds the identifiers of new workers to
the list of available workers. In this version of the system, this
method is called by the main block to configure a static deployment
scenario with a fixed number of workers. Methods \Abs{getWorker} and
\Abs{releaseWorker} move workers from the lists \Abs{available} to
\Abs{inuse} and vice versa. These methods are called by the service
endpoint to assign processing jobs to the workers. Method
\Abs{firingWorker} permanently removes a worker from the list of
available workers. Methods \Abs{getNbrAvailableW} and
\Abs{getNbrInuseW} calculates the number of available workers and the
number of workers that are in use, respectively. Note that we are not
using the last three methods in this version of the system, they are
used later in Section~\ref{sec:dynamicSys}.  Figure~\ref{fig:manager}
also contains the \ABS implementation of the service endpoint.  This
class has one method \Abs{invokeService} which is called by the
clients and forwards the client request to one of the workers.

      \begin{figure}[h!]
\begin{absexamplen}

interface LoadBalancer {
  Unit addWorker(Worker w);
  Worker getWorker();
  Unit releaseWorker(Worker w);
  Worker firingWorker();
  Int getNbrAvailableW();
  Int getNbrInuseW();
}

class RoundRobinLoadBalancer() implements LoadBalancer {
  
  List<Worker> available = Nil; List<Worker> inuse = Nil;

   Unit addWorker(Worker w){
       available = appendright(available,w);
     }
      
   Worker getWorker(){
       await (available != Nil);  
       Worker w = head(available); available = tail(available); inuse = appendright(inuse,w);  
       return w;
     }

   Unit releaseWorker(Worker w){
       available = appendright(available,w); inuse = without(inuse,w); 
      }

   Worker firingWorker(){
       await (available != Nil);
       Worker w = head(reverse(available)); available =  without(available,w);
       return w;
     }
     ...
}

interface SE { 
    Bool invokeService(Rat cost);
}

class ServiceEndpoint(LoadBalancer lb, Duration responseTime) implements SE {
    Bool invokeService(Rat cost){
        Time started = now();
        Worker w =  await lb!getWorker();
        Bool success = await w!process(cost,started,responseTime);
        await lb!releaseWorker(w);
        return success;
    }
 }
\end{absexamplen}
  \caption{\label{fig:manager} \ABS model of the load balancer and the service endpoint.}
\end{figure}

\paragraph{Static Deployment.}  Figure~\ref{fig:main1} models the
static deployment of the system as depicted by
Figure~\ref{fig:architecture} (without the autoscaler). In this case
we create the database, the load balancer and the service endpoint,
afterwards we deploy a fixed number of worker on the cloud. We also
need to include the workload that the system will process (as
described in Section~\ref{sec:environment}).

 \begin{figure}[h!]
\begin{absexamplen}
{    ...
    Database db = new Database();
    LoadBalancer lb = new RoundRobinLoadBalancer();
    SE endpoint = new ServiceEndpoint(lb, respTime);
    //start workers
    Int ctr = 0;
    while (ctr<nWorkers) {
       Fut<DC> fs =  cloud!launchInstance(map[Pair(Speed, nResources)]);
       DC vm = fs.get;
       [DC: vm] Worker w = new WorkerObject(db);
       lb!addWorker(w);
       ctr=ctr+1;
      }
    //start clients to generate a desired  workload
    ...
}
  \end{absexamplen}
  \caption{\label{fig:main1} Static deployment of the service.}
\end{figure}

\section{Extending the Service with Autoscaling and Dynamic Deployment}
\label{sec:dynamicSys}
In this section, we extend the model with an autoscaler to allow dynamic reconfiguration of the deployment on the cloud. 

Figure~\ref{fig:autoscaler} models an autoscaler which increases and
decreases the number of workers deployed on the cloud depending on the
number of available workers.  The \Abs{run} method models the initial
deployment of workers. The method \Abs{resize} acts as a monitor which
periodically checks if we need to adjust the number of workers. We
use the following \emph{ad hoc} policy: If the number of
available workers is less than one quarter of the total number of
workers, then we triple the number of available workers. If the number
of available workers is greater than one third of the busy workers,
then we reduce the number of available workers to half.

\begin{figure}[h!]
\begin{absexamplen}
class Autoscaler(CloudProvider cloud, LoadBalancer lb, Int nbrOfWorkers, Int nResources,
                       Database db, Int cycle) implements Autoscaler {

  Unit run(){
      Int ctr = 0;
      while (ctr<nbrOfWorkers) {
         Fut<DC> fs =  cloud!launchInstance(map[Pair(Speed, nResources)]);
         DC vm = fs.get;
         [DC: vm] Worker w = new WorkerObject(db);
         lb!addWorker(w);
         ctr=ctr+1;
      }
      this!resize();
    }
   
    Unit resize(){
      Int ctr = 0;
      await duration(cycle, cycle);
      Int available = await lb!getNbrAvailableW();
      Int inuse = await lb!getNbrInuseW();      
      if (available < ((available+inuse)/4)){ 
        ctr = 0;
        Rat extraworkers= 2*inuse;
        while (ctr<extraworkers) {
           Fut<DC> fs =  cloud!launchInstance(map[Pair(Speed, nResources)]);
           DC vm = fs.get;
           [DC: vm] Worker w = new WorkerObject(db);
           await lb!addWorker(w);
           ctr=ctr+1;
        }
      }
      if ((inuse/3 < available) && (available > nbrOfWorkers)){
        ctr = 0;
        Rat removeworkers= available/2;
        while (ctr<removeworkers) {
            Worker w = await lb!firingWorker();
            DC dc = await w!getDC();
            Bool down = await cloud!shutdownInstance(dc);
            ctr=ctr+1;
        }
     }
     this!resize();     
   }
 }
  \end{absexamplen}
  \caption{\label{fig:autoscaler} \ABS model of the autoscaler.}
\end{figure}

\paragraph{Dynamic Deployment.}  Figure~\ref{fig:main2} models the
initial deployment of the system as depicted in
Figure~\ref{fig:architecture}. In this case we create the database,
the load balancer, the autoscaler with an initial number of workers,
and the service endpoint. We also need the workload, as described in
Section~\ref{sec:environment}.

 \begin{figure}[h!]
\begin{absexamplen}
{    ...
    Database db = new Database();
    LoadBalancer lb = new RoundRobinLoadBalancer();
    Autoscaler as = new Autoscaler(cloud,lb,nWorkers,nResources,db,interval);
    SE endpoint = new ServiceEndpoint(lb, respTime);
    //start clients to generate a desired  workload
    ...
}
  \end{absexamplen}
\caption{\label{fig:main2} Dynamic deployment of the service.}
\end{figure}


\section{Comparing the Simulation Results}

Using the static and dynamic deployment models described above, we now analyze the behavior of our system and compare different scenarios using simulations.  Having early analysis results at design time allow us to observe how the model complies with certain non-functional properties capturing quality of service, and therefore related to SLAs. As an example, let us analyze if the model satisfies the following SLA, formulated in terms of model time:

\begin{center}
\begin{minipage}{0.95\linewidth}\em 
``The service must maintain a response time of less than 10 time intervals with an average success rate of at least 90\%. In addition, for an interval of 300 time intervals, the billing cost should not exceed the amount of 250000, where the billing cost is 50 per virtual machine, charged every 5 time intervals.''  
\end{minipage} 
\end{center}

\noindent
To assess possible deployments to meet  this SLA, we compare four scenarios. For static deployment, we consider three scenarios by varying the number of workers, and hence the number of virtual machines, between 80,~100, and 120. The fourth scenario is a dynamic deployment with the autoscaler.  We run each scenario 100 times using the \ABS simulator. In each run we record the total number of workers and the number of busy workers which have been in use in each time interval. We also record the success rate of the invocations made by the clients (with a deadline of 10 time intervals)  and the accumulated billing cost of running the scenarios until time 300. 
\begin{figure}[t!]
\centering
 \includegraphics[width=\linewidth]{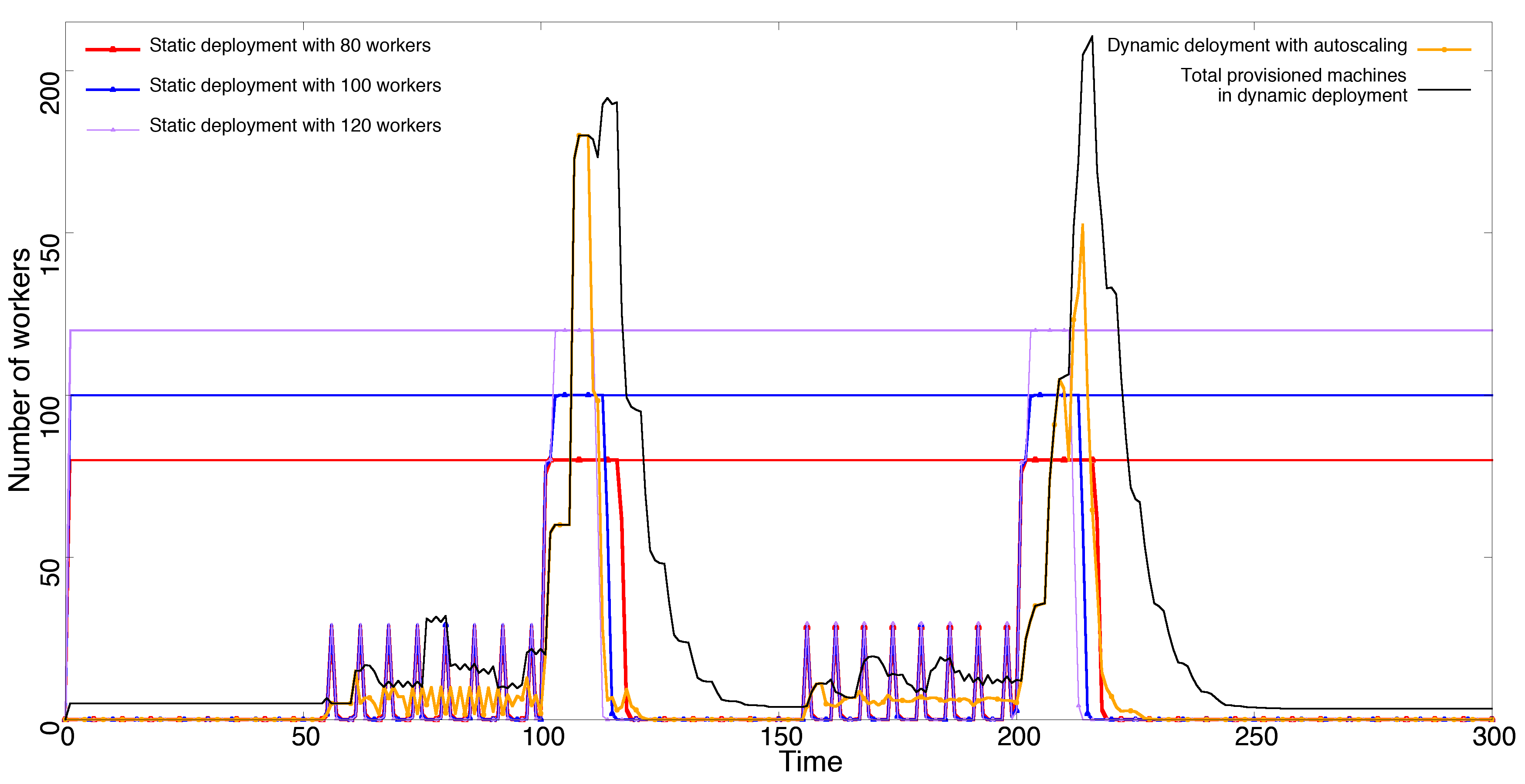}
  \caption{Average provisioning of virtual machines and machines in use for the different static and dynamic deployment scenarios}
  \label{fig:workers_in_use}
\end{figure}

\begin{figure}[h!]
\centering
\includegraphics[width=0.75\linewidth]{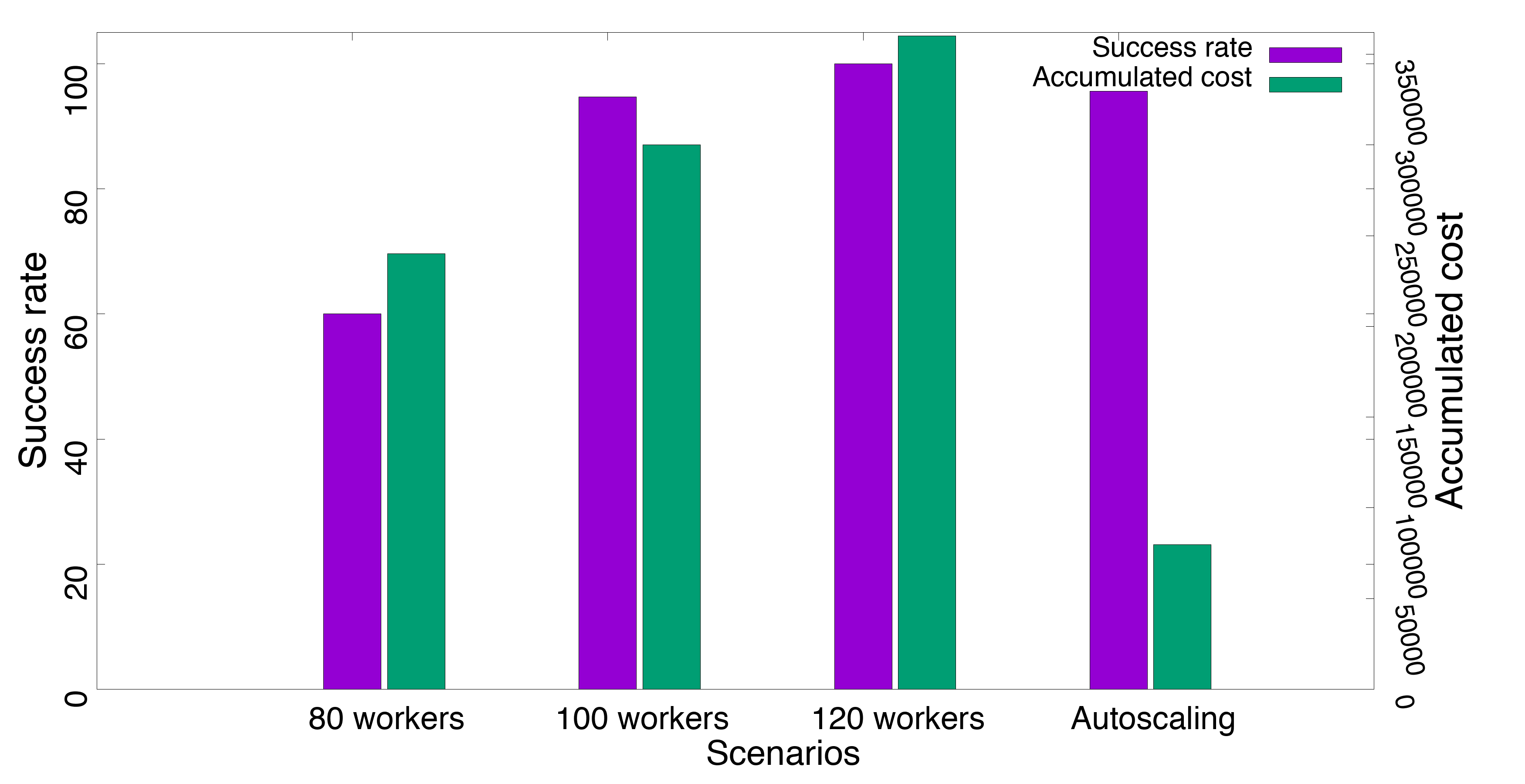}
  \caption{Average number of successful requests and total billing cost until time 300 for the different static and dynamic deployment scenarios}
  \label{fig:cost_quality}
\end{figure}

Figure~\ref{fig:workers_in_use} shows the number of provisioned machines (total number of workers) and the number of machines in use (workers in use) per time interval for each scenario, averaged over the 100 simulations. The red, blue, and purple lines in this figure capture the static deployment scenarios with 80, 100, and 120 workers, respectively. The straight vertical lines represent the statically fixed number of provisioned machines (virtual machines deployed on the cloud) in each time interval, while the oscillating lines capture the number of machines that are actually in use per time interval. In the case of the dynamic deployment scenario, the black and orange lines capture the varying number of provisioned machines and of machines in use per time interval, respectively.  For the static deployment scenarios, we can observe \emph{over-provisioning} of machines when the workload is low and \emph{congestion} when the workload has peaks (see Figures~\ref{fig:workload}~and~\ref{fig:workload2} for the description of the workload).  Note that the workload peaks occur when open clients are created, as indicated in Figure~\ref{fig:workload}. For the scenarios with static deployment, the period of congestion is inversely proportional to the total number of workers: the higher the number of workers, the shorter the period of congestion. For the dynamic deployment scenario, the total number of machines and the number of machines in use varies according to the workload, and the duration of both periods with over-provisioning of machines and with congestion are significantly shorter than for the static scenarios.

Figure~\ref{fig:cost_quality} compares the success rates (shown as the percentage of requests which have been executed within their deadline) and the accumulated billing costs for all scenarios until time 300, averaged over the 100 simulations. The success rate is scaled with the left y-axis while the cost is scaled with the right one.  In the static deployment scenarios, the success rate improves as the total number of machines increases. In fact the third scenario with 120 machines was chosen because it has 100\% success rate. For the dynamic deployment scenario, the average success rate, approximately 96\%,  is much better than for the static scenario with 80 machines (60\% success rate), and is close to the static scenario with 100 machines (95\% success rate) while it is a bit lower than the static scenario with 120 machines. When comparing the accumulated billing costs, the dynamic scenario is substantially lower than any of the static scenarios. In conclusion, the dynamic scenario offers the best trade-off between cost and quality of service which complies with the SLA described above.


\section{Concluding Remarks}
This paper revisits the main modeling concepts of \ABS for modeling
scalable and elastic services deployed on the cloud. We show by a
simple example how different deployment decisions can be expressed and
how these decisions affect the service-levels, formulated in terms of
the success rate of service requests, and the accumulated billing cost
for deploying the service, calculated from a pricing policy for the
virtual machines. The accuracy of predictions made using \ABS models
has been demonstrated on Hadoop benchmarks \cite{lin16fase} and on
industrial case studies \cite{ABHJSTW13}. We believe model-based
decision making about deployment strategies enable higher quality and
more cost-efficient services, both as part of the initial development
process and as a tool for DevOps teams.

\bibliographystyle{eptcs}
\bibliography{refs}
\end{document}